\documentclass[twocolumn,showpacs,preprintnumbers,amsmath,amssymb,prc]{revtex4}
\usepackage{graphicx}   
\graphicspath{{./}{./figs/}}
\begin{document}
%
%
\title{Experimental spectra analysis in THM with the help of simulation based on Geant4 framework}
\author{Chengbo Li}
\thanks{Supported by National Natural Science Foundation of China (11075218, 10575132) and Beijing Natural Science Foundation (1122017)}
\email{licb2008@gmail.com}
\affiliation{Beijing Radiation Center,  Beijing 100875, China}
\affiliation{Key Laboratory of Beam Technology and Material Modification of Ministry of Education,College of Nuclear Science and Technology, Beijing Normal University, Beijing 100875, China}

\author{Qungang Wen}
\email{qungang@ahu.edu.cn}
\affiliation{Anhui University,  Hefei 230601, China}

\author{Shuhua Zhou}
\author{Yuanyong Fu}
\author{Jing Zhou}
\author{Qiuying Meng}
\affiliation{China Institute of Atomic Energy, Beijing 102413, China}

\author{Zongjun Jiang}
\author{Xiaolian Wang}
\affiliation{State Key Laboratory of Particle Detection and Electronics, USTC, Hefei 230026, China}

%
%

%

%
\date{\today}
\begin{abstract}
The Coulomb barrier and electron screening cause difficulties in directly measuring nuclear reaction cross sections of charged particles in astrophysical energies. 
The Trojan-horse method has been introduced to solve the difficulties as a powerful indirect tool. 
In order to understand experimental spectra better, Geant4 is employed to simulate the method for the first time. 
Validity and reliability of the simulation are examined by comparing the experimental data with simulated results.
The Geant4 simulation can give useful information to understand the experimental spectra better in data analysis and is beneficial to the design for future related experiments.

\end{abstract}

\maketitle
\section{Introduction}

Understanding energy production and nucleosynthesis in stars requires increasingly precise knowledge of the nuclear reaction rates at the energies of interest \cite{sun}.
However, at astrophysical temperature, nucleus react at very low energies much lower than the Coulomb barrier for charged particles. 
The reaction cross sections are very small due to the Coulomb barrier, so that the direct measurement is almost imposable. 
To overcome the experimental difficulties arising from the small cross sections and the electron screening, the Trojan-horse method (THM) \cite{gbaur, stypel, cspitaleri, thmdist, allthm, cheng, sromano, wen1, wen2} has been introduced. 

THM provides a valid alternative approach to measure unscreened low-energy cross sections of reactions between charged particles.
In the method, suitable three body reactions are measured under the quasi-free kinematic conditions with beam energies above their Coulomb barrier. 
The method can also be used to retrieve information on the electron screening potential when ultra-low energy direct measurements are available.

Geant4 \cite{geant41, geant42} is a toolkit for the simulation of the passage of particles through matter. 
Its  application areas include high energy, nuclear and accelerator physics, as well as studies in medical and space science. 

In this paper, for the first time, we develop a simulation program based on the Geant4 framework for THM research in order to understand the experimental spectra better.

\section{Trojan Horse Method}

The Trojan-horse method belongs to an indirect measurement method in experimental nuclear astrophysics.
The basic assumptions of the THM have been discussed extensively elsewhere \cite{sun, gbaur, stypel, cspitaleri, allthm} and detailed theoretical derivation of the formalism employed can be found in \cite{stypel} . 

\begin{figure}
\begin{center}
\includegraphics[width = 0.30\textwidth]{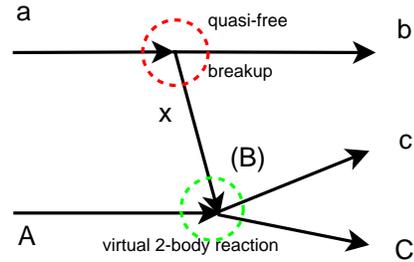}\\
\caption{Diagram of Trojan-horse method}
\label{fig1} 
\end{center}
\end{figure}

The diagram of THM is shown in Figure \ref{fig1}.
The method is based on quasi-free (QF) reaction mechanism, which allows us to derive indirectly the cross section of a two-body reaction
\begin{equation}\label{eq:two_body}
    A+x \rightarrow C+c
\end{equation}
from measurement of a suitable three-body process
\begin{equation}\label{eq:three_body}
    A+a \rightarrow C+c+b
\end{equation}
The nucleus $a$ is considered to be dominantly composed of clusters $x$ and $b$ ( $a=x \oplus b $).

After the breakup of nucleus $a$ due to the interaction with nucleus $A$, the two-body reaction occurs between the transferred particle $x$ and nucleus $A$ whereas nucleus $b$ does not participate and acts as a spectator. 
The energy in the entrance channel $E_{Aa}$ is chosen above the height of the Coulomb barrier, so as to avoid a reduction in cross section. 

At the same time, the effective energy of the reaction between $A$ and $x$ can be relatively small, mainly because the energy $E_{Aa}$ is partially used to overcome the binding energy 
$\varepsilon_a$ of $x$ inside $a$  (Eq.(\ref{eq:Eqf1})), and the Fermi motion of $x$ inside $a$ compensates at least partially for the $A+a$ relative motion (Eq.(\ref{eq:Eqf2})) . 
\begin{equation}\label{eq:Eqf1}
        E_{Ax}^{qf}=E_{Aa}\left(1-\frac{\mu_{Aa}}{\mu_{Bb}}\frac{\mu_{bx}^{2}}{m_x^2}\right)-\varepsilon_{a}
\end{equation}
\begin{equation}\label{eq:Eqf2}
        E_{Ax}=E_{Ax}^{qf}\pm E_{xb}
\end{equation}

Since the transferred particle $x$ is hidden inside the nucleus $a$ (so called Trojan-horse nucleus) and the collision of $A$ with $x$ takes place in the nuclear interaction region, the two-body reaction is free of Coulomb suppression and, at the same time, not affected by electron screening effects.

Thus the interesting two-body reaction cross section can be extracted from the measured three-body reaction using the relation formulation Eq.(\ref{eq:sec3all}) after selecting the quasi-free events:
\begin{equation}\label{eq:sec3all}
\frac{d^3\sigma}{dE_{Cc}d\Omega_{Bb}d\Omega_{Cc}} = KF  |W|^2  P_l\frac{d\sigma_l}{d\Omega}({Ax\rightarrow Cc})
\end{equation}
where KF is the kinematical factor,  $W$ is the momentum distribution of the spectator $b$ inside the Trojan-horse nuclei $a$,   and $P_l$ is the penetration function

In our work, the THM have been used to study two important astrophysical nuclear reactions related with $^{9}Be$ abundance.
\begin{equation}\label{eq:aLi6}
    ^2H(^9Be,\alpha  ^6Li)n \Longrightarrow ^9Be(p,  \alpha ) ^6Li
\end{equation}
and
\begin{equation}\label{eq:dBe8}
    ^2H(^9Be,d ^8Be)n \Longrightarrow  ^9Be(p, d ) ^8Be
\end{equation}
where deuteron is used as the Trojan horse nucleus, due to its $d=p \oplus n$ structure \cite{ thmdist} ,  the proton acts as a participant while the neutron is a spectator to the virtual two-body reaction.

\section{Experiment setup}

\begin{figure}
\begin{center}
\includegraphics[width = 0.45\textwidth]{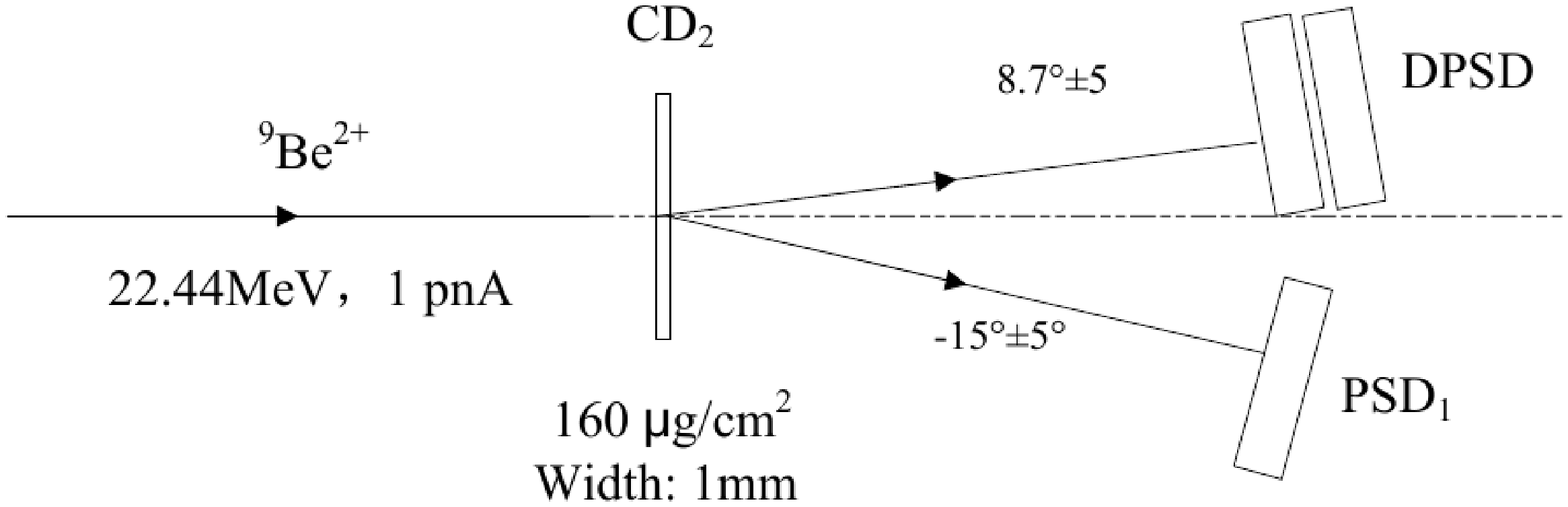}\\
\caption{Experiment setup of Trojan-horse method for the reaction Eq.(\ref{eq:dBe8})}
\label{fig2} 
\end{center}
\end{figure}

The measurements of the reactions Eq.(\ref{eq:aLi6}) and Eq.(\ref{eq:dBe8}) were both performed in Beijing National Tandem Accelerator Laboratory at China Institute of Atomic Energy. 
The experimental setup for the reaction Eq.(\ref{eq:dBe8}) was installed in the nuclear reaction chamber at the R60 beam line terminal as shown in Figure \ref{fig2}. 
A $\rm ^{9}Be^{2+}$ beam at 22.44 MeV provided by the HI-13 tandem accelerator was used to bombard a deuterated polyethylene target $\rm CD_{2}$ placed vertically to the beam axis. 
The thickness of the target is about $\rm 160 \mu g/cm^{2}$. In order to reduce the angle uncertainty coming from the large beam spot, a linear target with 1 mm width was used.

A position sensitive detector ($\rm PSD_1$) was placed at $15^{\circ} \pm 5^{\circ}$ to the beam line direction and about 240 mm from the target to detect outgoing deuterons, 
and a DPSD (Dual Position Sensitive Detector, consisted of $\rm PSD_u$ in the upside and $\rm PSD_d$ downside ) was used at $8.7^{\circ} \pm 5^{\circ}$  in the other side of the beam line and 250 mm distance from the target to detect two alpha particles decayed from the unstable outgoing particle $\rm ^{8}Be$.
The arrangement of the experimental setup was modelled in Monte Carlo simulation in order to cover a region of quasi-free angle pairs.
The trigger for the event acquisition was given by coincidence of signals from the PSD and DPSD.

The reactions Eq.(\ref{eq:aLi6}) can also be measured with $\rm PSD_1$ detecting alpha and $\rm PSD_u$ detecting  $\rm ^{6}Li$ particles in coincidence. 


\section{Experimental spectra analysis with the help of Geant4 simulation}

The first step of the data analysis work is the energy and angle calibration of PSD and DPSD. 
After the calibration of the detectors, we have the energe and momentum of the particles detected by $\rm PSD_1$, $\rm PSD_u$ and $\rm PSD_d$.
Then we reconstructed $\rm ^{8}Be$ from ($\rm E_u$, $\rm E_d$, $\rm \theta_u$, $\rm \theta_d$) on the assumption that the particles detected by DPSD are two $\alpha$.
The energy and momentum of the third particle $n$ of the exit channel  $\rm ^9Be + d \rightarrow ^8Be + d + n$  are calculated from ($\rm E_1$, $\rm E_2$, $\rm \theta_1$, $\rm \theta_2$), where particle1 is  $d$ and particle2 is  $\rm ^8Be$.

The most important thing to do before using THM to extract information of the 2-body reaction $\rm^{9}Be+p\rightarrow^{8}Be+d$ from the 3-body reaction $\rm ^9Be + d \rightarrow ^8Be + d + n$  is to select the right events which satisfied with the three body reaction of quasi-free reaction mechanism apart from all the other outgoing channels.
There are many exit channels from the same entrance channel of $\rm^{9}Be+d$, for example, the $\rm^{9}Be+^{2}H\to\alpha+^{6}Li+n$ channel can be detected as well.
Other than the exit channels from $\rm^{9}Be+d$, there are more other outgoing channels from the reaction of the beam bombard to other elements in the target such as $\rm ^{12}C$ and $\rm ^{1}H$. 

Therefore, it is particularly important to understand the experimental spectrum in the events selections.
In order to understand the experimental spectrum better, Geant4 simulation is applied to the THM study in our work. 

Geant4 \cite{geant41} \cite{geant42} developed by CERN is a well established Monte Carlo framework for simulation of particles passage through matter. 
Detector and target construction parameters in Geant4 simulation program of THM were defined according to the experiment setup. And the default $\rm FTFP\_BERT$ physics list was used in the process. 
An event generator code was written in $\rm C^{++}$ to create momentum information of outgoing particles from different nuclear reactions.

Some of the Geant4 simulation results will be shown below comparing with the experimental data.

\subsection{$\rm E-\theta$ spectrum of 2-body reactions}

%

\begin{figure}[h]
\begin{center}
\begin{minipage}[c]{0.22\textwidth}
\includegraphics[width =\textwidth]{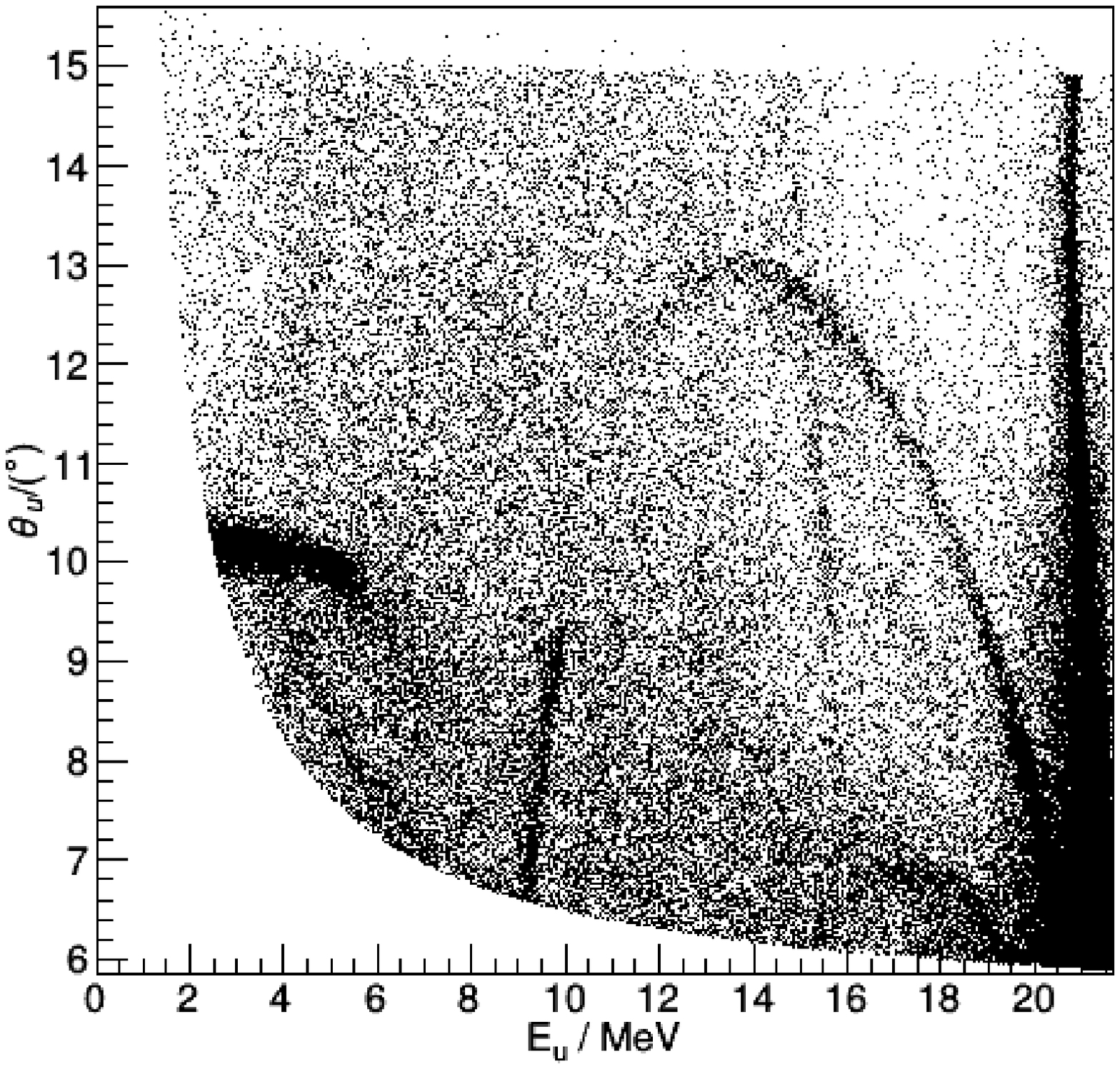}
\end{minipage}
\begin{minipage}[c]{0.22\textwidth}
\includegraphics[width =\textwidth]{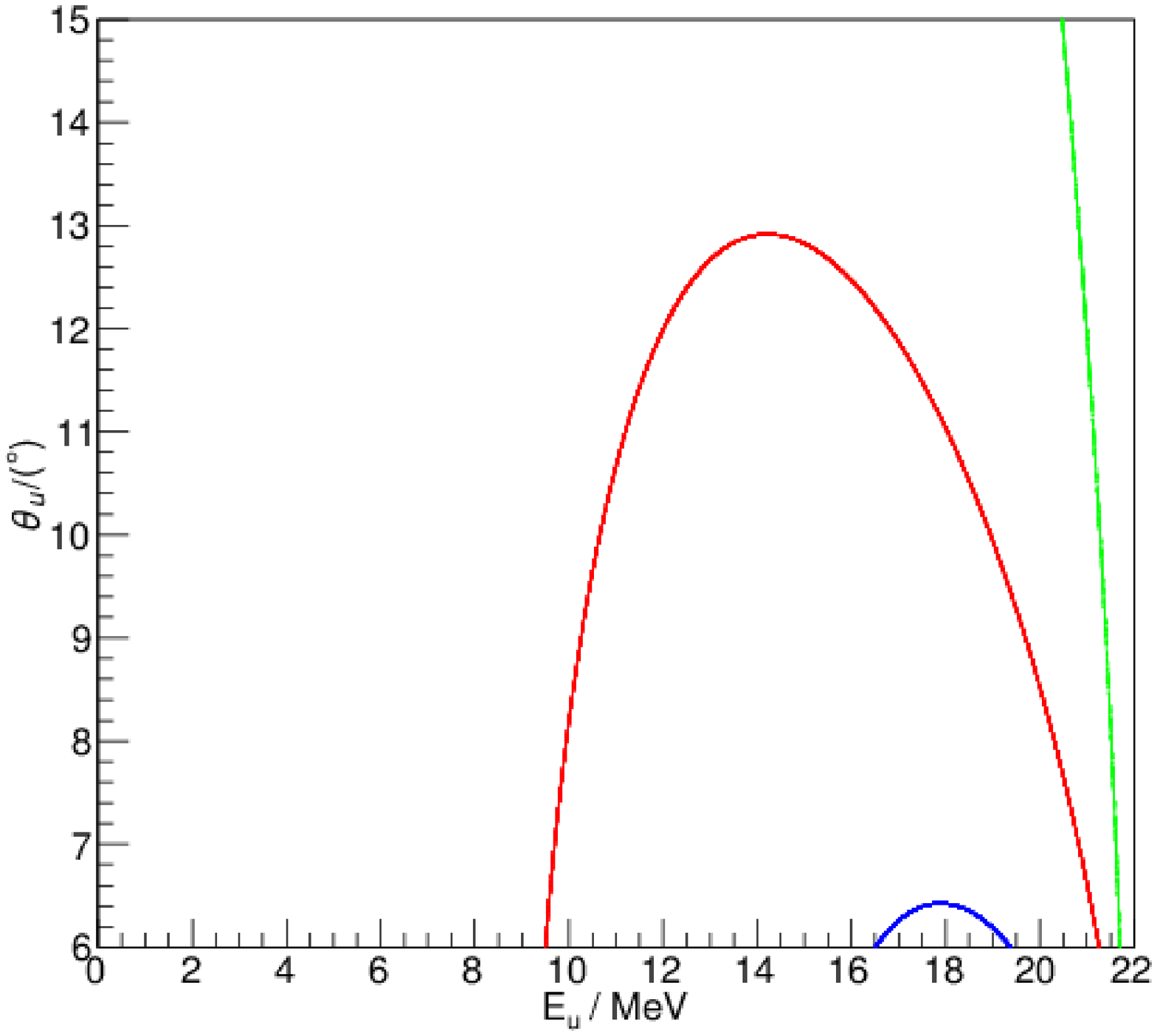}
\end{minipage}
\begin{minipage}[t]{0.45\textwidth}
\caption{Comparison of experimental spectrum $\rm E_{u}-\theta_{u}$ (left) with simulated one (right) }
\label{fig_vs1} 
\end{minipage}
\end{center}
\end{figure}

Figure \ref{fig_vs1} (left) shows the experimental spectrum of $\rm E_{u} - \theta_{u}$ detected by $\rm PSD_u$. 
The red points in Figure \ref{fig_vs1} (right) shows Geant4 simulation of the reaction $\rm^{9}Be+^{2}H\to^{9}Be+^{2}H$. 
The $\rm E_{^9Be}-\theta_{^9Be}$ curve looked like parabola is easy to find in the experiment spectrum. 

There is also a small arc between $\rm E_{u}(16MeV-20MeV)$ in Figure \ref{fig_vs1} (left). 
The simulation result shows that it comes from the $\rm^{9}Be+^{1}H\to^{9}Be+^{1}H$ elastic scattering process (the bule points in Figure \ref{fig_vs1} (right) ).This means that there are also some $\rm ^1H$ in the $\rm CD_2$ target.

The simulation result of the $\rm^{9}Be+^{12}C\to^{9}Be+^{12}C$ elastic scattering process is also shown in Figure \ref{fig_vs1} (right, the green points), which meets the curve of $\rm E_{u}\sim 22 MeV$ in the experimental spectrum.

The $\rm E-\theta$ curve of outgoing particles from two body reactions can give a validity test to the detector calibration.  
It can also give information of elements in target.

\subsection{$\rm E_1-E_u$ spectrum: kinematic focus}

%

\begin{figure}[h]
\begin{center}
\begin{minipage}[c]{0.22\textwidth}
\includegraphics[width =\textwidth]{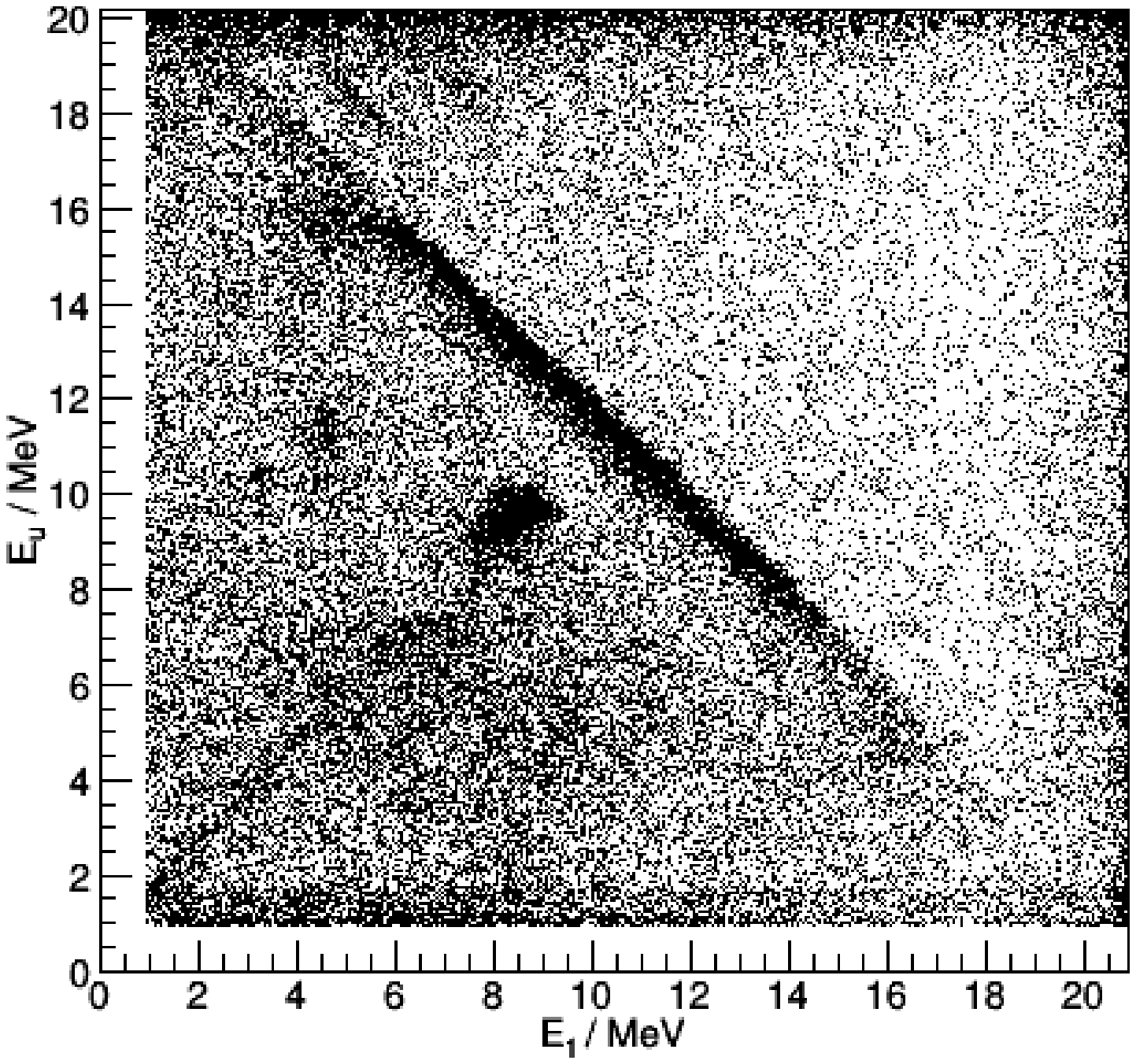}
\end{minipage}
\begin{minipage}[c]{0.22\textwidth}
\includegraphics[width =\textwidth]{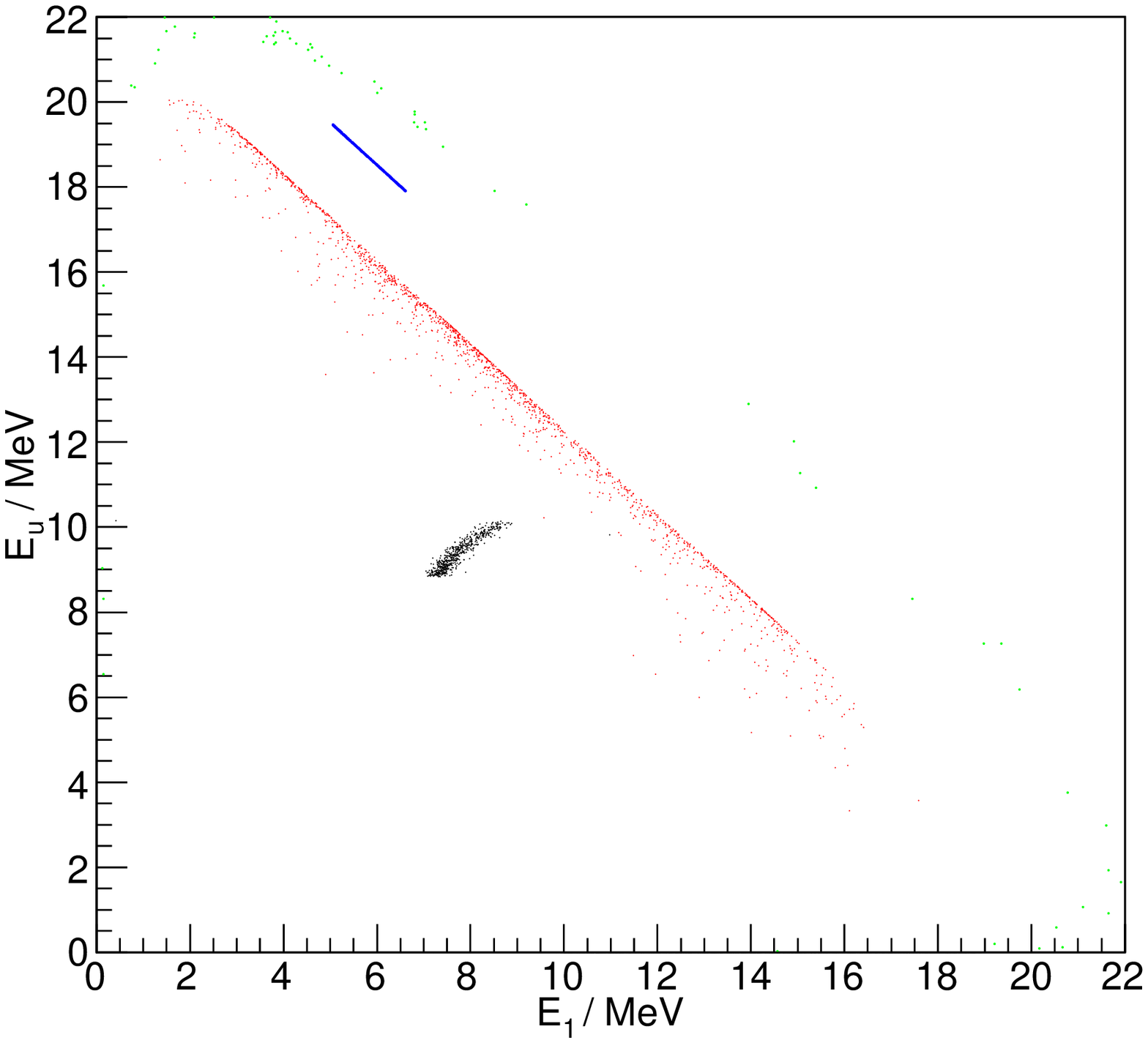}
\end{minipage}
\begin{minipage}[t]{0.45\textwidth}
\caption{Comparison of experimental spectrum $\rm E_{1}-E_{u}$ (left) with simulated one (right) }
\label{fig_vs2}  
\end{minipage}
\end{center}
\end{figure}

Figure \ref{fig_vs2} (left) is the two-dimensional energy spectrum $\rm E_{1}-E_{u}$ of experimental data. 
Simulation results of different reaction channels are shown in Figure \ref{fig_vs2} (right). 

The red points are the simulation of $\rm^{9}Be+^{2}H\to\alpha+^{6}Li+n$ reaction from quasi free process, which are interesting for THM reaction Eq.(\ref{eq:aLi6}).

The bule points are the simulation  of $\rm^{9}Be+^{1}H\to\alpha+^{6}Li$ reaction caused by the beam bombarding to $\rm ^1H$ in the target.

The green points are the simulation of $\rm^{9}Be+^{2}H\to^{3}H+^{4}He+^{4}He$ reaction channel, which are not easy to find out.

All these points can be found in the experiment spectrum.

The black points, which puzzled us for a long time, are the simulation result of $\rm^{9}Be+^{2}H\to^{9}Be+^{2}H$ elastic scattering results. Normally, the spots of the elastic scattering in the two-dimensional energy spectrum should be in the line of $\rm E_{1}+E_{u}=22.4 MeV$. We finally found out the reason by the simulation program. It is because the detector can not deplete all the energy of the emitted high energy deuteron particles due to the limitation of the detector thickness (500$\rm \mu m$). 

It can be seen that the simulation program can give us great help to get a better understanding to the experiment spectrum.

\subsection{$\rm E_{u}-E_{d}$ spectrum: reconstruction of $\rm^{8}Be$ }

%

\begin{figure}[h]
\begin{center}
\begin{minipage}[c]{0.22\textwidth}
\includegraphics[width =\textwidth]{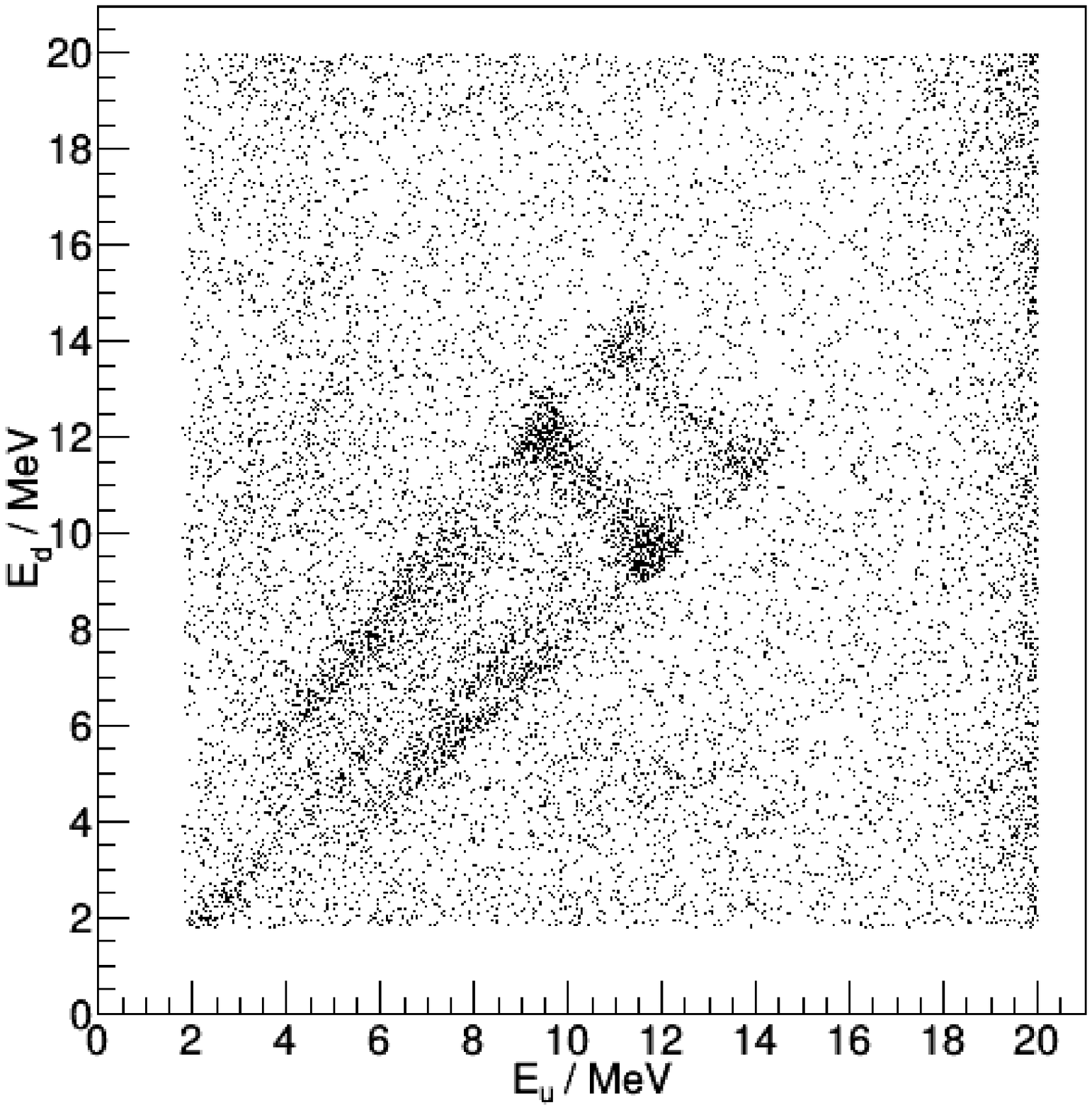}
\end{minipage}
\begin{minipage}[c]{0.22\textwidth}
\includegraphics[width =\textwidth]{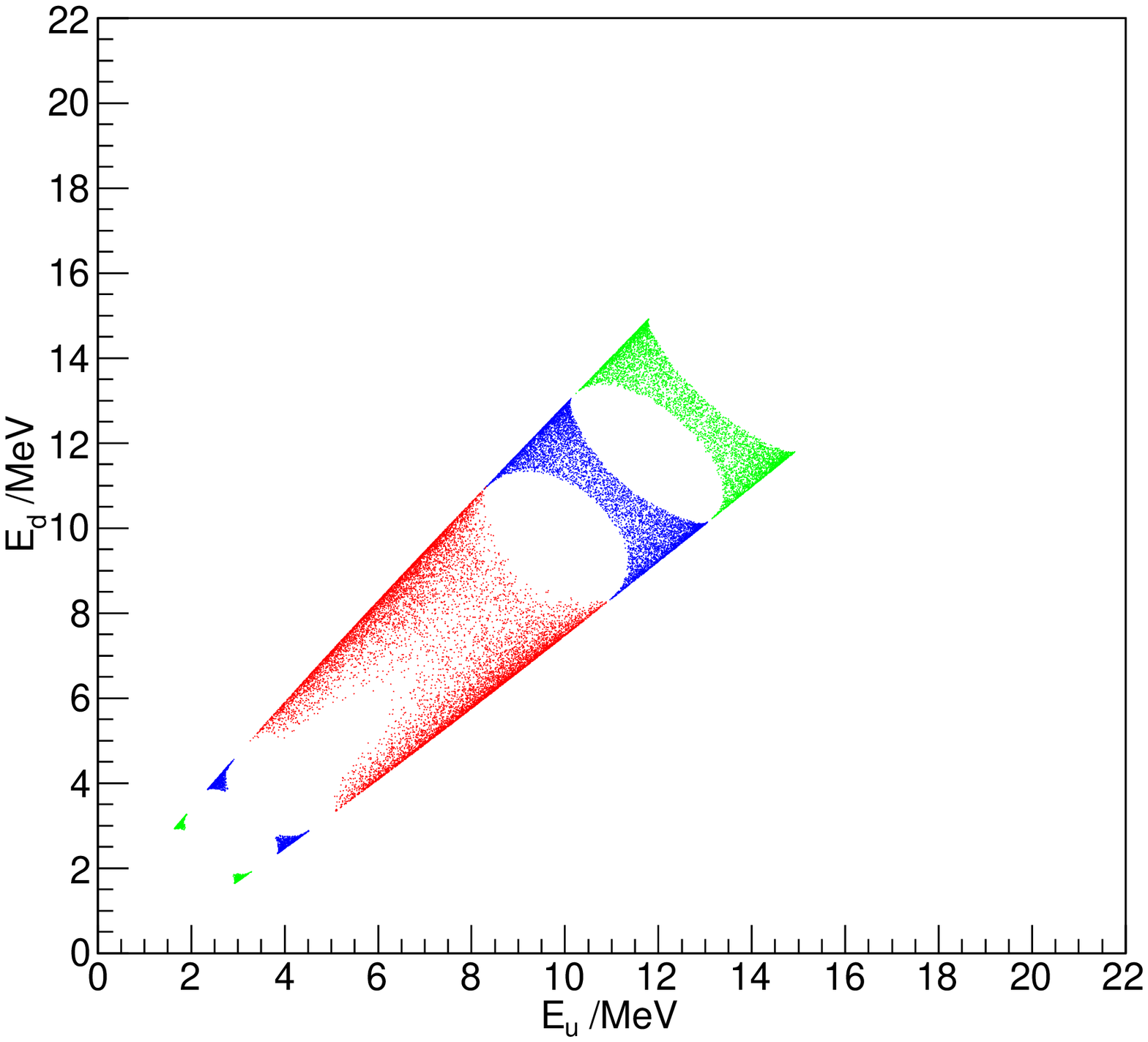}
\end{minipage}
\begin{minipage}[t]{0.45\textwidth}
\caption{Comparison of experimental spectrum of reconstruction for $\rm^{8}Be$ (left) with simulated one (right) }
\label{fig_vs3} 
\end{minipage}
\end{center}
\end{figure}

The important step in data analysis is the reconstruction of $\rm ^{8}Be$ particle from two $\alpha$ particles detected by DPSD.
Figure \ref{fig_vs3} (left)  is the experimental spectrum of $\rm E_{u}- E_{d}$.

The simulation result of $\alpha$ particles decayed from $\rm ^{8}Be$ of  $\rm^{9}Be+^{2}H\to^{8}Be+^{2}H+n$ reaction is shown in Figure \ref{fig_vs3}  (right, the red points). 
You can see the agreement between the simulated data and the low energy range of the experimental data ($\rm E_{u}\in (5 MeV-10 MeV)$).
And we find from the simulation that the high energe parts are not from the $\rm^{9}Be+^{2}H\to^{8}Be+^{2}H+n$ reaction channel which is interested for us.

A simulation of $\alpha$ particles decayed from $\rm ^{8}Be$ of  $\rm^{9}Be+^{2}H\to^{8}Be+^{3}H$ reaction channel is given in Figure \ref{fig_vs3}  (right, the green points). 
It is in good agreement with the experimental data points located on the high energy area ($\rm E_{u}\in (12 MeV-14 MeV)$). 

Another simulation of $\alpha$ particles decayed from  $\rm ^{8}Be$ ground state of $\rm ^{9}Be+d \rightarrow ^{8}Be^{*}+ ^{3}H \rightarrow ^{8}Be+^{3}H+\gamma$ reaction channel after the $\rm ^{8}Be^{*}$ transfered from the first excited state to the ground state by emitting a gamma ray is shown in Figure \ref{fig_vs3}  (right, the blue points), 
which can show good agreement with the middle energy range ($\rm E_{u}\in (10 MeV-12 MeV)$) in the experimental data.

With the help of the simulation, we can find out the origin of different parts of the experiment data, thus we can choose the the right events by a graphical cut only including the interesting reaction channel in the experimental spectrum to avoid the interference of other channels.

\subsection{Future applications}
The simulation code is also very useful in the research work such as the energy loss and angle dispersion of the particles passing through a $\rm \Delta E$ detector or the dead layer of detectors.
The simulation results can help us in the design of the THM experiment, as well as the particle identification and error analysis in data analysis.



\section{Summary}

A simulation system based on the Geant4 framework was established and applied to the Trojan horse method experimental study for the first time.  
The validity and reliability of the simulation system are examined by comparing the experimental data with the simulated results in our work.
The simulation system can provide useful information to understand the experimental spectra better in data analysis, an it is  beneficial to the design for future related experiments.

\begin{acknowledgments}
We thank Dr. Chengjian Lin and Dr. Xia Li from CIAE for their kind help during the experiment measurement. 
We also thank Prof. C. Spitaleri and his research group from INFN-LNS laboratory of Italy for the precious collaboration in the THM study.
In addtion, we thank Dr. Zhiyi Liu for his kind discussion in the paper writting.
\end{acknowledgments}

\end{document}